**Valerio Arnaboldi, Andrea Passarella, Marco Conti**

Institute of Informatics and Telematics (IIT), National Research Council (CNR), Via G. Moruzzi 1, 56124, Pisa, Italy

**Robin Dunbar**

Department of Experimental Psychology, University of Oxford, South Parks Road Oxford, OX1 3UD, United Kingdom

Department of Information and Computer Science, Aalto University School of Science, Konemiehentie 2 - FI-02150, Espoo, Finland

## Abstract

We analyze the ego-alter Twitter networks of ~300 Italian MPs and 18 European leaders, and of about 14,000 generic users. We find structural properties typical of social environments, meaning that Twitter activity is controlled by constraints that are similar to those shaping conventional social relationships. However, the evolution of ego-alter ties is very dynamic, which suggests that they are not entirely used for social interaction, but for public signaling and self-promotion. From this standpoint, the behavior of EU leaders is much more evident, while Italian MPs are in between them and generic users. We find that politicians – more than generic users – create relationships as a side effect of tweeting on discussion topics, rather than by contacting specific alters.

1. **Structure of Ego-alter relationships of Politicians in Twitter**

Many politicians, especially in the US and in Europe, are nowadays adopting Online Social Networks (OSN), and Twitter in particular, as an official channel to communicate with their constituency and the wider public in general, and to maintain strategic relationships with other politicians (Jackson & Lilleker, 2011), to the point of impacting major political events, such as the last three US presidential elections[1] (e.g., Stirland, 2008; Greengard, 2009; Conway, Kenski, and Wang, 2013).

Even though these previous studies identified differences in the use of Twitter by politicians compared to generic users, none of them, to the best of our knowledge, considered, in detail, the analysis of "ego-alter ties", i.e., the set of social relationships that a given user (the ego) maintains with other social peers (alters) – as defined in Section 2.2 below. Dunbar et al. (2015) found that OSN ego-alter ties have the same structural properties as 'offline' ego-alter ties (those maintained through face-to-face interactions). This indicates that the use of OSN, including Twitter, may not substantially change the size and pattern of our social relationships, which are instead limited by the "social cognitive capacity" of the human brain (Dunbar, 1993).

In this paper, we analyze specifically the ego-alter networks of a representative set of politicians (a large fraction of Italian MPs and the most important European leaders on Twitter) . It is known already that the Twitter relationships characteristic of politicians are qualitatively different than those of generic users, as the former are primarily meant for self-promotion and public signaling, and not for establishing social support (Murthy 2012, Enli and Simonsen 2017). However, politicians still invest time and cognitive resources to maintain such relationships. Thus, we hypothesize that they could be constrained by the same limits that shape relationships established for social support. Therefore, a goal of this paper is to quantitatively analyze whether politicians' ego-alter networks show similar features to generic users' networks, despite the different purpose of relationships for the two classes. Specifically, we analyzed the ego-alter networks of about 300 Italian MPs and 18 EU leaders, and compared them with those of about 14,000 generic users, who are neither politicians nor public figures of any kind. Note that, our analysis is data-driven, in the sense that it is not based on detailed models and parameters about politicians' ego-alter networks. Thus, the structural model we derive is entirely driven by data analysis. While this approach might have drawbacks, we think it minimizes the risks of results being biased by specific modelling assumptions.

---

[1] https://www.wired.com/2016/11/facebook-won-trump-election-not-just-fake-news/

We found that, when observed over long time intervals, the ego-alter networks of politicians in Twitter show a structure similar to those found in other social environments, both online and offline. On the other hand, when observed over shorter time intervals, these ego networks are not stable over time: the online relationships of politicians in Twitter come and go with a very high turnover, particularly for *high-frequency* relationships. This feature appears also for generic users, and was previously observed in other studies on Twitter (Arnaboldi, Conti, Passarella, & Dunbar, 2013). In offline social networks, high-frequency relationships usually correspond to stable and intimate relationships (Hill & Dunbar, 2003). Therefore, this confirms that Twitter relationships are of a somewhat different nature when observed one-by-one. However, our results also show, to the best of our knowledge for the first time, that when considering the entire set of ego-alter relationships of any given ego (politician or not), it exhibits constraints very similar to those observed, in general, in offline social networks. Moreover, we hypothesize that, for politicians in particular, online relationships are more related to the content that is exchanged between users and exposed to the public than by the need to exchange social support. We find that, for the analyzed politicians, a very significant percentage of relationships is created as a by-product of discussions on topics (coded in tweets with *hashtags*), and much more so than is the case for generic users. Finally, a regression analysis shows that the intensity and diversity of use of hashtags explains quite well the frequency of interaction over the relationships, particularly for politicians.

## 2. Data and Methods

### 2.1. Direct Tweets and Dataset description

Ego-alter ties are usually characterized by their strength (i.e., the intensity or importance of relationships). In this work, we use the frequency of *direct tweets* between users to measure this. Direct tweets can be performed through *mentions* (user names can be included in a tweet, with the effect that the related users will receive the tweet in their home pages), *replies* (a mechanism that allows users to reply to tweets by others and engage in conversations with them) and *retweets* (re-share of tweets created by others).

We collected the tweets created by a set of Twitter accounts of politicians with a sufficient level of direct communication activity. Users whose engagement in the platform is too sporadic are not relevant for understanding how ego networks are created and maintained over time, and have been excluded. We considered accounts for which we can access at least six months of Twitter communication. Previous research (Arnaboldi et al., 2013; Wilson et al., 2012) demonstrated that Twitter relationships can be considered stable only after six months of activity, because only after that does the rate of new relationships created over time remain constant. In addition, we discarded accounts with a sporadic use of the platform, i.e., with periods of short duration of very high activity and long periods of inactivity. To do so, we used the following heuristic. We looked at the distribution of the number of direct tweets generated by each account every month of its activity, and we selected only the users who sent at least one direct tweet every three days (as an average frequency within each month) for at least 50% of the total number of months of their activity.

We collected two samples from Twitter, and filtered them as described above. The first (hereinafter "sample 1: Italian MPs") includes 304 members of the Italian Government Cabinet[2] and members of the Italian Parliament[3]. The second sample ("sample 2: EU Leaders") contains 18 European Leaders[4]. We accessed the data from Twitter between June and December 2015, using the Twitter REST API and obtaining the complete history of the tweets generated by users, up to the limit of 3,200 tweets per user imposed by the API. The list of European leaders contains, among others, the accounts of David Cameron and Matteo Renzi, two of the most followed EU leaders on Twitter as of 2014[5]. The complete list of European leaders in sample 2 is reported in Table 1, together with their overall tweet frequency.

We also collected the accounts of 100,000 randomly selected Twitter users, filtered as explained above. In addition, from these accounts we automatically removed those that are not managed by human agents, which are not used for socializing with other people (e.g., bots, spammers, companies), and accounts of public figures (e.g., famous actors or political leaders). To this end, we used an account classifier built on a support vector machine (SVM) – a standard machine learning method that, after being trained on a set of manually classified accounts (the *training* set), is able to automatically classify unknown accounts. The same classifier has been used also by Arnaboldi et al. (2013), and has an accuracy greater than 80%. After this filtering phase, we obtained 13,728 accounts of generic Twitter users. This is referred hereafter as "sample 3: generic users sample".

---

[2] obtained from the Italian Government official website: http://www.governo.it/Governo/Ministeri/ministri_gov.html

[3] obtained from a website collecting open data about the Italian Parliament: http://parlamento17.openpolis.it/

[4] obtained after filtering, as described above, a public Twitter list of European leaders' accounts, excluding the accounts not associated with individuals: https://twitter.com/twiplomacy/lists/european-leaders/members

[5] https://marcorecorder.com/2014/03/07/personal-twitter-accounts-of-european-prime-ministers-whos-scoring-best/

### 2.2. Methods for Ego Network Analysis

#### 2.2.1. Ego Network model

Several definitions of ego network have been proposed in the literature, depending on the focus of the study. In sociology, an ego network generally consists of a single actor (ego) together with the actors the ego is connected to (alters) and all the links among those alters (Everett & Borgatti, 2005). Another possible model considers only alter-alter ties, discards ego-alter ties, and is used to analyze the topological features of the social context in which the ego is immersed – e.g., (Burt, 2001; McCarty, 2002). In this paper, we focus instead only on ego-alter ties, and use the Dunbar ego network model (Hill & Dunbar, 2003; Sutcliffe, A., Wang, D., Dunbar, R.I.M., 2012). Through this model (hereinafter the 'ego network model'), it is possible to characterize human cognitive constraints and how they impact key social processes such as cooperation and willingness to share resources (Sutcliffe, et al., 2012).

Figure 1 depicts the typical representation of a Dunbar ego network (Sutcliffe, Wang & Dunbar, 2012). Two key aspects of human sociality are analyzed through this type of ego networks. The first is the social capacity of individuals, measured as the number of alters whom an ego can *actively* maintain over time in the network. Active relationships are defined as those that involve at least one interaction per year, which is interpreted as a signal of a minimal level of cognitive investment (Hill & Dunbar, 2003). The number of active relationships is known as the *Dunbar's Number* and is around 150 both in offline and online social networks (Dunbar, 1993; Dunbar et al., 2015). The second aspect is the hierarchical organization of alters into concentric social circles around the ego (Zhou et al., 2005; Hill & Dunbar, 2003). These circles are hierarchically inclusive: each circle includes all the alters of its inner circles, whereas *rings* refer to the portions of circles that exclude inner circles. Inner circles include stronger relationships (typically alters from whom the ego seeks strong social support) while outer circles include acquaintances. The typical average sizes of circles (Roberts et al., 2009; Dunbar et al., 2015) are 1.5, 5, 15, 50, and 150. The ratio between adjacent circles is around 3, and this is considered a key aspect of human ego networks (Zhou et al., 2005). It is also known that these structures do not change much over time (Saramäki et al., 2014), and in particular can also be found in popular online social networks, i.e., Facebook and Twitter (Dunbar et al., 2015).

### 2.2.2. Twitter ego networks: indices for static and dynamic analysis

From the three direct communication mechanisms of Twitter (i.e. replies, mentions, and retweets), we built, for each user obtained after the filtering process described in Section 2.1, three ego networks. As shown in previous work (Dunbar et al., 2015), it is possible to find the layers of ego networks through cluster analysis (Bishop, 2006) on the frequencies of contact. Cluster analysis automatically divides the ego-alter relationships into groups with homogeneous values of contact frequency, each group representing a ring. One of the simplest of such techniques is *k*-means, a method that divides the data into a predefined number of clusters, *k*. The optimal *k* for a set of data points can be found using various standard metrics, such as the Akaike Information Criterion (Bishop, 2006). The optimal *k* found in Twitter is typically five (Dunbar et al., 2015). Thus, we used cluster analysis on our samples, and applied *k*-means with *k*=5 on each ego network.

To analyze the dynamic properties of ego networks, we divided the time series of interactions into time windows of one year each, and computed the ego networks for each time window using the same clustering method described above, as was previously done by Arnaboldi et al. (2013). To analyze the evolution of ego network structure over time, month by month, we considered the following 11 months of communications, obtaining a series of one-year time windows, with an 11-months overlap between consecutive elements in the series. On the other hand, to analyze the turnover rate within the ego networks and the stability over time of the ego network rings, we considered non-overlapping consecutive time windows. Specifically, for each ring, we analyzed its stability in each pair of adjacent time windows, and averaged over all adjacent time windows. We chose a temporal length of one year so as to be able to capture relationships with at least one message a year, which coincide with the definition of active network in the literature (Roberts et al., 2009). For this part of the analysis, we considered all three types of direct tweets together, as the goal was to characterize overall dynamic behavior.

### *Network structure stability indices*

We used two indices to measure the stability of the ego network rings over time. First, we calculated the *Jaccard index*, which, given two sets of elements (which, in our case, are two adjacent time windows containing their respective alters), is defined as the cardinality of the set intersection divided by the cardinality of the set union, and is thus equal to 1 when the two sets are identical, and 0 when they have no elements in common. Second, we defined the *Jump index*, $C$, which counts the number of movements, or jumps, between rings (averaged over all alters in a given ring), with each movement between adjacent rings counting as one jump, and a movement from ring $x$ (with values between 1 and 5, where 1 is the innermost ring and 5 the outermost one) to outside the ego network (or vice versa) as $5 - x + 1$ jumps. Note that we discarded all cases where no movement occurs, i.e., the index is the jump conditioned to moving to a different ring. We also defined a normalized version of the index by dividing each number of jumps for a given alter by the maximum possible number of jumps that alter could make from its ring in the first time window.

### *Indices for the use of hashtags over Twitter relationships*

We also analyzed the relation between the use of hashtags and the dynamic properties of ego networks. For this reason, we separated the relationships that are "activated" by hashtags (i.e., which show a hashtag in the first contact associated with them) from the other relationships; we then analyzed (i) the proportion, (ii) the frequency of contact and (iii) the set of hashtags used for the two types of relationships, as a function of the layers of the ego networks in which they lie. Finally, we studied the relation between the contact frequency of Twitter direct relationships and a series of indices that quantify the *intensity* and the *diversity* in the use of hashtags on those relationships in all our datasets (described in detail in Section 3.5).

### 3. Results

### 3.1. Use of Twitter direct communication by Politicians

Table 1 shows the average Tweet frequency of EU leaders and Italian MPs (#of tweets per day). In general, this is compatible with a human use of the platform. Specifically, leaders show a higher Twitter activity, which confirms prior studies on the lack of equalization among politicians of this OSN (Van Aelst et al., 2017). The percentages of direct tweets usage (for each type of direct tweet) for the European leaders and Italian MPs are reported in Table 1. We note that most EU Leaders show a marked preference for one type of direct communication in Twitter (though which type is preferred varies between politicians). This is shown by the ratio between the first and the second highest percentages per user (*F-S* ratio in the table), which is typically much higher than 1, and also high on average for all the EU Leaders ($M = 3.48$) as well as for Italian MPs ($M = 3.08$). The average percentages show that EU leaders slightly prefer mentions (46% over 42% retweets), while Italian MPs prefer retweets (43% over 36% mentions). Retweets are typically a sign of endorsement of the original tweet, and it might be expected that it is more often used by non-leaders than leaders in a social group.

### 3.2. Static Properties of Twitter ego Networks of Politicians

Table 2 reports the size of the identified circles and the scaling ratio between adjacent circles. The table is divided into three sections, each of which is dedicated to one specific type of direct communication. Each section reports the properties of the Dunbar ego-network circles of All Italian MPs (sample 1), EU Leaders (sample 2), and the subset of Italian MPs and EU Leaders who share the specified preferred type of direct communication. For individual EU Leaders, we note a significant variability in the sizes of the layers. In some cases, the sizes of the layers are quite close to the values in the Dunbar ego network model (e.g., Matteo Renzi for replies, Donald Tusk for mentions, Charles Michel for retweets). In other cases, some layers, particularly the external ones, may be either much smaller or much bigger than those in the ego-network model. Such a marked variability is common in offline ego networks (Hill and Dunbar, 2003; Roberts, et al. 2009, Zhou et al. 2015). In our analysis, this is expected for two reasons. On the one hand, sizes in the ego-network model are *average* values derived across a population, and all datasets analyzed in the literature show a significant variability across individuals. Moreover, Twitter is only one possible means of interaction, which does not exhaust the total interaction capacity of individuals, and individual leaders may use them differentially. This is also confirmed by a positive correlation between the size of ego networks and the tweeting activity of politicians ($r = 0.39$, $p < .01$ for replies, $r = 0.5$, $p < .01$ for mentions, $r = 0.47$, $p < .01$ for retweets). On the other hand, as expected, when the results are averaged over all the ego networks in the samples (All Italian MPs and All EU Leaders), and over each subset of Italian

MPs with a specific preferred type of direct communication, the sizes of the circles are similar to those of the ego-network model shown in Fig. 1.

Analyses of the scaling ratio provide even more insights. The scaling ratio between adjacent circles averages, for Italian MPs, 2.6 for replies, 2.5 for mentions and 3.0 for retweets and, for EU Leaders, 2.2 for replies, 2.3 for mentions, and 2.7 for retweets. Aside from the impressive consistency of these ratios (and there is no empirical reason why this should be so), these values are close to the value of 3 for the Dunbar ego-network model. To compare the variability of sizes and scaling ratios with each other, we defined index $C$, which is the ratio between the width of the confidence interval of the mean and the mean itself. The average value of $C$ for the *sizes* of the circles is 0.16 for replies, 0.10 for mentions, and 0.13 for retweets for Italian MPs, and 1.42 for replies, 0.55 for mentions, and 0.55 for retweets for EU Leaders. The value of $C$ for the *scaling ratios* is 0.08 for replies, 0.07 for mentions, and 0.07 for retweets, for Italian MPs, and 0.39 for replies, 0.25 for mentions, and 0.33 for retweets, for EU Leaders. This shows that, even though the sizes of layers can vary considerably across politicians, the scaling ratios vary a lot less. This suggests that the structural properties of the ego networks are remarkably constant across users, resulting in scaling ratios that are quite stable.

### 3.3. Dynamic Ego Network Properties

Overall, we found a marked dynamism over time in the use of Twitter, particularly for EU Leaders. As an example, Figure 2 shows the number of non-direct and direct tweets per month for Matteo Renzi, David Cameron, Jean-Claude Juncker, and Andrzej Duda. Vertical lines show the most important political events in the countries of the specific leader. Tweeting activity varies significantly over time, and is particularly high before key political events. This result agrees with findings in the literature on legislative ambition: politicians who seek higher offices usually try to enlarge the size of their electorate community, changing their public behavior to appeal to as many people as possible (Black, 1972).

Figure 3 depicts the values of the Jaccard index and the Jump index (raw and normalized), averaged for all the ego networks for politicians and for generic users. The results indicate that there is a very high variability in the inner layers of the ego networks for both politicians and generic users, whereas the outer rings are progressively more stable. This tells us that Twitter ego networks are substantially different from the ego networks analyzed in environments such as phone call networks, where inner layers are found to be more stable than outer ones (Saramäki et al., 2014). In addition, the high values of the Jump index for the inner layers indicate that alters tend to enter these layers from external parts of the ego network *directly* (the mean number of jumps from or to *R1* is around 3 over a maximum number of 5 jumps), and they return to these external regions when they quit the inner rings. On the other hand, alters in *R5* tend to jump to the adjacent rings (the mean number of jumps is close to 1) when they move. This means that looking at ego-networks over one-year time windows (dynamic ego networks), inner layers are much less stable than the outer ones.

It is interesting to understand the relationship between layers in the static and dynamic ego networks. Specifically, Figure 4 shows a separate plot for each layer in the static network. For ego-alter ties in a given static layer, it shows the average frequency of appearance in the layers of the dynamic ego networks. Note that the "OUT" layer groups all ties outside the active network. For generic users, the figures show that, ego-alter ties of a given layer in the static network appear with the highest frequency in the same layer of the dynamic ego networks. However, for inner static layers (R1-R4), the distribution of average frequencies inside dynamic layers is more dispersed than for outer static layers (R5-OUT). In addition, for any static layer, even for the most internal ones, there is a significant frequency of appearance in the OUT dynamic layer, i.e., outside the ego network. It is also interesting to note that Italian MPs show a behavior similar to the generic users, while for EU Leaders there is a less marked correspondence between the static and the dynamic views. This agrees with the high burstiness found in Twitter communications for EU Leaders, also shown in Figure 2, which results in ephemeral high-frequency relationships.

### 3.4. The use of hashtags over Twitter relationships

We conjecture that, especially for politicians, the findings we reported in the preceding section may reflect the use of Twitter as a tool to discuss – and make the public aware of – a range of different topics through intense interactions with alters. This is based on the observation that, for politicians in particular, Twitter is not only used to broadcast political messages but also to engage conversation on political opinions (Tumasjan et al. 2010). To explore this further, we first analyzed whether ego-alter ties are created as a "by-product" of exchange of information on specific topics, and whether this is more evident for the politicians than for generic users. We also analyzed whether there is a significant difference in the two classes of users in the number of hashtags exchanged over ego-alter ties. Finally, we analyzed whether intensity and diversity of use of hashtags can explain the frequency of contact over ego-alter ties, again for both classes of users.

We measured the number of new alters and new hashtags added in the ego network every month, considering "time 0" for each ego network as the point in time when the ego activated its first relationship. Their average values are 9.78 and 14.13 for Italian MPs, and 11.24 and 0.67 for generic users, respectively[6]. The number of new alters contacted by politicians and generic users have the same order of magnitude, whereas for the new hashtags the value is much higher for Italian MPs.

---

[6] We omit the results for EU Leaders since the confidence intervals for their means are too large to be significant.

We then analyzed whether there is an evidence of ego-alter ties being created *because* of the use of hashtags, and, if this is the case, the pattern of interactions over these relationships. We separated relationships "activated" by hashtags (i.e., containing a hashtag in the first direct tweet over that relationship) from all the other relationships. The analysis shows a significant difference between politicians and generic users. The relationships activated by hashtags are 15% of the total number of relationships for EU Leaders, 17% for Italian MPs, and 5.9% for generic users. Figure 5 shows more detailed statistics. The graphs in the first column indicate that, for Italian MPs and EU Leaders, the percentage of relationships activated by hashtags is higher than for generic users in all ego network layers. Graphs in the second columns show that, in contrast to generic users, the mean frequency of contact of relationships activated by hashtags for Italian MPs and EU Leaders is similar to that of other relationships. Finally, graphs in the third and fourth columns show that the number of hashtags exchanged (both in total and for distinct hashtags) is higher for politicians (both Italian MPs and EU Leaders). In addition, relationships activated by hashtags show higher values than relationships not activated by hashtags in all layers for Italian MPs and EU Leaders, whereas there is no such clear trend in the ego networks of generic users.

All in all, these results indicate that (i) politicians tend to activate ego-alter ties *because* of the topic discussed and do so *more* than generic users do, (ii) they maintain a higher frequency of interactions, compared to generic users, over those ties, with respect to the frequency over the rest of the ties, (iii) in general they use many more hashtags than generic users, and (iv) they do so preferentially over ties activated by hashtags, much more than generic users do. All this is consistent with the use, particularly by politicians, of Twitter ego-alter links as a means to expose to the public positions about specific topics, *through an interaction* with a specific alter.

### 3.5. Relation between contact frequency and use of hashtags

Finally, we analyzed the relationship between the indices defined to capture the intensity and diversity in the use of hashtags and contact frequency, through regression analysis on a ring-by-ring basis. As a first measure of *intensity* of hashtag use over a relationship $r$, we calculated, for relationships activated by a hashtag $h_{act}$, the number of direct tweets containing that hashtag $N(r, h_{act})$. We also calculated this measure cumulatively for the ego $e$, as the number of times $h$ has been used by the ego in all its relationships, $N(e, h_{act})$. These measures embody the importance for the users of the hashtags that activated their relationships. The indexes are clearly null for relationships not activated by any hashtags. Indices $N(r, h_{max})$ and $N(e, h_{max})$ are analogous to the previous measures, but are related to the hashtag ($h_{max}$) that appears with the highest frequency in relationship $r$, instead of the activating hashtag. Finally, to measure the *diversity* of the hashtags in direct tweets, we calculated $D_{rel}(r)$ (the number of hashtags appearing in a relationship $r$) and $U_{rel}(r)$ (the number of distinct hashtags appearing in $r$).

The results of the regression analysis are reported in Table 3. We report the values of $R^2$ for all the different models created. $R^2$ is the coefficient of determination of the models, and measures the degree to which the model is able to approximate contact frequency (values range from 0 – worst fit – to 1 – best fit). We built different linear regression models for EU Leaders, Italian MPs, and generic users, considering only relationships activated by hashtags or other relationships separately, and considering all the relationships in the ego networks together or dividing them into the different circles. All the estimates for the regression coefficients show positive signs and are consistent among the models. It is worth noting that, in general, the $R^2$ values are medium/high for most models – $R^2$ higher than 0.3 is usually considered a sufficiently high value. Moreover, values are higher for politicians than for generic users, and are higher for relationships activated by hashtags than for the other relationships. This confirms the hypothesis that direct contact frequency in Twitter is largely correlated with the intensity and diversity of use of hashtags, i.e. that the strength of ego-alter ties is explained by the need to expose communication on a topic or person to the public, and that this is even more marked for politicians.

## 4. Discussions and Conclusion

For EU Leaders and Italian MPs, the structural patterns of ego network layers are very close to those found in the literature for offline social networks. It is worth noting that the *scaling ratio* between the adjacent layers is always close to three, which is the typical value found in Dunbar ego networks. This is an indication that, when controlling for Twitter activity, online ego networks may be shaped by the same cognitive and time constraints at the basis of the structure of offline social networks. This tells us that, although politicians' Twitter ego-alter ties are largely a means of self-promotion (Enli and Simonsen 2017), the way they organize their Twitter contacts obey the same principles that underpin the formation and management of human social relationships in general, which are related to cognitive constraints imposed by the human brain.

The analysis of the dynamic evolution of ego networks over time revealed that the structural properties of politicians' ego networks in our samples (and also those of generic Twitter users) are rather unstable over time, especially in their inner layers (on a year-by-year basis, 80% turnover in the inner layers vs. 40% turnover in the outer layers), which was unexpected. If Twitter is used to maintain social relationships exactly as any other communication means, one would expect more stable relationships in inner than in outer layers, since in social environments contact frequency is a proxy of tie strength and emotional closeness. This suggests that the nature of ego-alter ties in Twitter is significantly different than in offline social environments, in particular for Italian MPs and EU Leaders. The high turnover also matches a bursty use of Twitter that we observed for many politicians and the EU leaders in particular: sudden increments in the use of direct tweets coincide with important political events where the leader seeks higher office or higher popular support.

A notable difference between politicians and generic users is in the way politicians use ego-alter ties to contribute to public discussions, to advertise positions on certain topics, or for self-promotion. Politicians have a much higher percentage of relationships "activated" by hashtags (i.e., showing a hashtag in their first direct tweet) than generic users in all ego network layers. Moreover, their activity (frequency of contact) on these relationships is comparable with respect to their activity on other relationships, while generic users are much less active on relationships activated by hashtags. Moreover, the number of hashtags exchanged over relationships by politicians is much higher than for generic users, and it is markedly biased towards links activated by hashtags. This confirms that the use of Twitter relationships by politicians is a way to expose the public to particular information, rather than to keep social contacts with alters. Evidence about the higher use of hashtags by politicians has been found before (Enli and Simonsen 2017). However, we found that this does not apply just to general tweeting activity, but also to "social" tweets containing explicit mentions to other users. This is a sign of the mix between a social-centric and information-centric use of Twitter by politicians.

A regression analysis between the intensity and diversity of use of hashtags in direct tweets and the contact frequency of direct relationships provides further insights into the use of Twitter as an information-oriented platform, for both classes of analyzed users. Specifically, the intensity and diversity of hashtag use is positively correlated with the frequency of contact. This suggests that, as with generic Twitter users, politicians use Twitter as a means to discuss and exchange information publicly *through* direct relationships, rather than to maintain social relationships. For politicians, this behavior is even more evident than for generic users. Moreover, when analyzed on a ring-by-ring basis, models have better fit for inner layers than for outer layers, showing that intensity and diversity of use of hashtags are associated with high-frequency interactions.

Considering both (i) the high turnover of inner layers and (ii) the relationships between direct contact frequency and use of hashtags, we arrive at a view of *Twitter* use *as a mix of social- and self-promotion platforms* for generic users, but especially so for the politicians. For the latter, high contact frequency relationships involve a large number of topics (high *diversity*) with high frequency per topic (high *intensity*), but do not mean stable relationships with specific individuals (high *turnover*).

By and large, our results confirm general prior findings about the use of Twitter in the political sphere. Specifically, there is increasing evidence in the literature about Twitter being an important tool that politicians use in their public activities (Jungherr 2016). Starting from a relatively low use a few years ago (Larsson & Moe, 2010, Effing et al., 2011), politicians are increasingly using Twitter during electoral campaigns, with a noticeable effect on the electoral outcomes (Effing et al. 2011 and references herein; Stirland, 2008; Greengard, 2009; Conway, Kenski, and Wang, 2013). Also, the use of Twitter by politicians is changing. In addition to being a means for self-promotion and broadcasting (Enli and Simonsen 2017), it is increasingly (though not primarily (Jungherr 2016)) becoming a means for bidirectional interactions with voters, for mobilising supporters, for consulting (Tumasjan et al., 2010, Graham et al., 2013). Despite clear evidence about these facts, the literature highlights that a more detailed understanding on how politicians use Twitter is needed (Jungherr 2016), in particular through longitudinal studies on the use of Twitter by politicians (Graham et al., 2013).

We believe that our results go some way to filling this gap. First, they confirm previous findings about the increasing importance of the use of Twitter for politicians. Second, through the detailed analysis of the ego-alter structures provided in the paper, they could suggest ways to optimise political communication over Twitter. Our results show that, despite Twitter not being used as a tool for establishing social relationships, the same constraints observed in other social networks (both online and offline) control the behavior of the *entire set* of ego-alter ties for any given Twitter user, and this is true even for politicians (EU Leaders and Italian MPs). As the cognitive and time resources politicians can allocate to Twitter activities is limited, our results could suggest ways politicians should distribute their "Twitter time" over their ego-alter networks so as to achieve an optimally efficient allocation of their limited resources when engaging in this social platform. As it is known that social ties at the different levels of the Dunbar's ego network model results in different willingness of social peers to share resources and collaborate with the ego (Roberts et al. 2009), politicians may monitor the various levels of their ego networks to understand which alters are more likely to contribute spreading their messages, and tune their communication campaigns accordingly. For example, as it has been found (Arnaboldi et al., 2016) that trusted information flows preferentially through strong ties (due to their associated level of trust), politicians might want to engage Twitter influencers in their inner layers, to maximise the impact of their communications. Based on our results, an effective means to increase tie strength could be the use of hashtag relationships (thus including other users in discussions marked by specific topics), as witnessed by the higher values for communication frequency on relationships activated by hashtags. Thus, the use of hashtags could be useful for politicians both to consolidate interactions and foster information diffusion at the same time.

**Tables**

Structure of Ego-alter relationships of Politicians in Twitter                                                       23Table 1

*Social Tweets Usage by European Leaders and Italian MPs.*

| Name | % of social | % of Replies | % of Mention | % of Retweets | F-S ratio | Tweet Frequency |
|---|---|---|---|---|---|---|
| Matteo Renzi | 57.35 | **46.34** | 24.83 | 28.83 | 1.60 | 2.42 |
| David Cameron | 34.04 | 1.98 | **87.43** | 10.59 | 8.26 | 1.89 |
| Enda Kenny | 58.74 | .00 | **67.70** | 32.30 | 2.10 | .74 |
| Erna Solberg | 89.68 | **79.66** | 9.74 | 10.60 | 7.52 | 2.16 |
| Miro Cerar | 53.05 | 12.13 | 31.59 | **56.28** | 1.78 | 3.06 |
| Jean-Claude Juncker | 62.29 | 3.33 | 41.60 | **55.07** | 1.32 | 1.88 |
| Donald Tusk | 31.03 | 5.31 | **84.45** | 10.24 | 8.25 | 1.94 |
| Andrzej Duda | 81.67 | 46.59 | 4.26 | **49.15** | 1.05 | 4.15 |
| Alexis Tsipras | 61.76 | 2.17 | **77.96** | 19.87 | 3.92 | 2.81 |
| Taavi Rõivas | 79.09 | 7.57 | 5.99 | **86.44** | 11.42 | 5.85 |
| Nicos Anastasiades | 35.20 | .24 | 37.75 | **62.01** | 1.64 | 1.56 |
| Toomas Hendrik Ilves | 82.73 | 6.83 | **52.92** | 40.25 | 1.31 | 6.85 |
| Laimdota Straujuma | 64.94 | 17.21 | 20.71 | **62.08** | 3.00 | .95 |
| Borut Pahor | 32.66 | 13.26 | **59.77** | 26.97 | 2.22 | 4.04 |
| Atifete Jahjaga | 51.96 | 1.29 | 35.40 | **63.31** | 1.79 | .95 |
| Charles Michel | 74.44 | 17.63 | 24.29 | **58.08** | 2.39 | 1.69 |
| Pablo Iglesias | 65.41 | 7.48 | 42.50 | **50.02** | 1.18 | 7.60 |
| Pedro Sánchez | 68.68 | 1.02 | 45.09 | **53.89** | 1.20 | 6.67 |
| Mean for EU Leaders | 60.26 | 12.07 | **46.12** | 41.81 | 3.48 | 3.18 |
| Mean for Italian MPs (sample 1)[a] | 64.67 | 21.05 | 36.22 | **42.73** | 3.08 | 2.64 |

*Note.* Values representing the highest percentages for each user are in boldface. F-S ratio is the ratio between the first and the second highest percentages for the user. Tweet frequency is measured in number of tweets (both plain and social) created per day.

[a] For comparison with the rest of the politicians in the dataset.

Table 2

*Structural properties of the Twitter ego networks of politicians.*

| User | Measure | C1 | C2 | C3 | C4 | C5 |
|---|---|---|---|---|---|---|
| | | | *Reply Networks* | | | |
| All Italian MPs | Size (M±SD) | 1.58±.21 | 4.89±.70 | 12.3±.1.9 | 29.50±4.9 | 70.26±13.0 |
| | Ratio (M±SD) | | 3.27±.32 | 2.57±.20 | 2.41±.18 | 2.25±.18 |
| All EU Leaders | Size (M±SD) | 1.87±.9 | 4.51±4.1 | 11.86±10 | 26.52±20 | 47.43±31 |
| | Ratio (M±SD) | | 2.61±.89 | 2.06±.96 | 2.19±1.01 | 1.86±.08 |
| Matteo Renzi | Size | 1 | 6 | 15 | 48 | 107 |
| | Sc. ratio | | 6 | 2.5 | 3.2 | 2.2 |
| Erna Solberg | Size | 13 | 41 | 92 | 180 | 302 |
| | Sc. ratio | | 3.2 | 2.2 | 2.0 | 1.7 |
| Italian MPs who mostly use replies | Size | 2.07±.89 | 7.5±3.1 | 20.1±7.7 | 52±18.41 | 141±46.19 |
| | Sc. ratio | | 3.92±.95 | 2.87±0.5 | 2.84±.49 | 2.85±.48 |
| | | | *Mention Networks* | | | |
| All Italian MPs | Size (M±SD) | 1.27±.1 | 3.18±.28 | 6.86±.69 | 15.83±1.7 | 41.74±5.3 |
| | Ratio (M±SD) | | 2.62±.20 | 2.21±.15 | 2.35±.15 | 2.65±.22 |
| All EU Leaders | Size (M±SD) | 1.47±1.03 | 3.79±2.54 | 8.26±5.65 | 17.6±12 | 38.89±31 |
| | Ratio (M±SD) | | 2.71±0.97 | 2.24±0.8 | 2.14±0.75 | 1.97±0.77 |
| David Cameron | Size | 2 | 5 | 15 | 38 | 85 |
| | Sc. ratio | | 2.5 | 3 | 2.5 | 2.2 |
| Enda Kenny | Size | 1 | 5 | 15 | 28 | 39 |
| | Sc. ratio | | 5 | 3 | 1.9 | 1.4 |
| Donald Tusk | Size | 2 | 4 | 14 | 46 | 111 |
| | Sc. ratio | | 2 | 3.5 | 3.3 | 2.4 |
| Alexis Tsipras | Size | 1 | 2 | 6 | 12 | 57 |
| | Sc. ratio | | 2 | 3 | 2 | 4.8 |
| Toomas Hendrik Ilves | Size | 1 | 3 | 10 | 45 | 159 |
| | Sc. ratio | | 3 | 3.3 | 4.5 | 3.5 |
| Borut Pahor | Size | 3 | 9 | 18 | 33 | 78 |
| | Sc. ratio | | 3 | 2 | 1.8 | 2.4 |
| Italian MPs who mostly use mentions | Size | 1.32±.18 | 3.41±.54 | 7.7±1.39 | 19.53±3.4 | 58.9±10 |
| | Sc. ratio | | 2.7±.37 | 2.32±.26 | 2.63±.27 | 3.17±.36 |
| | | | *Retweet Networks* | | | |
| All Italian MPs | Size (M±SD) | 1.59±.20 | 4.82±.60 | 12.84±1.6 | 35.25±4.5 | 118.2±17.1 |
| | Ratio (M±SD) | | 3.25±.3 | 2.74±.2 | 2.86±.2 | 3.3±.26 |
| All EU Leaders | Size (M±SD) | 1.14±.35 | 3.08±1.66 | 7.51±6.16 | 19.59±20 | 72.02±70 |
| | Ratio (M±SD) | | 2.72±1.18 | 2.32±0.89 | 2.46±0.94 | 3.18±2.39 |
| Miro Cerar | Size | 1 | 3 | 5 | 15 | 59 |
| | Sc. ratio | | 3 | 1.7 | 3 | 3.9 |
| Jean-Claude Juncker | Size | 1 | 2 | 5 | 13 | 46 |
| | Sc. ratio | | 2 | 2.5 | 2.6 | 3.5 |
| Andrzej Duda | Size | 1 | 6 | 26 | 101 | 296 |
| | Sc. ratio | | 6 | 4.3 | 3.9 | 2.9 |
| Taavi Rõivas | Size | 1 | 2 | 6 | 23 | 269 |
| | Sc. ratio | | 2 | 3 | 3.8 | 11.7 |
| Nicos Anastastades | Size | 1 | 3 | 4 | 11 | 29 |
| | Sc. ratio | | 3 | 1.3 | 2.8 | 2.6 |
| Laimdota Straujuma | Size | 2 | 4 | 7 | 16 | 39 |
| | Sc. ratio | | 2 | 1.8 | 2.3 | 2.4 |
| Atifete Jahjaga | Size | 2 | 4 | 8 | 14 | 40 |
| | Sc. ratio | | 2 | 2 | 4.25 | 2.41 |
| Charles Michel | Size | 2 | 8 | 23 | 41 | 119 |
| | Sc. ratio | | 4 | 2.9 | 1.8 | 2.9 |
| Italian MPs who mostly use retweets | Size | 1.63±.28 | 5.13±.92 | 14.23±2.41 | 39.81±6.3 | 134.9±22.8 |
| | Sc. ratio | | 3.36±.42 | 2.88±.26 | 3.02±.27 | 3.52±.36 |

Structure of Ego-alter relationships of Politicians in Twitter 25Table 3

*Coefficient of determination ($R^2$) of different regression models created using measures of the use rate of hashtags by Twitter users to explain the contact frequency of their social ties.*

|  | Links with an activating hashtag | | | | | | Links w/o an activating hashtag | | | | | |
|---|---|---|---|---|---|---|---|---|---|---|---|---|
|  | Linear Regression | | | | | | | | | | | |
| Rings | ALL | R1 | R2 | R3 | R4 | R5 | ALL | R1 | R2 | R3 | R4 | R5 |
| Italian MPs | .68 | .56 | .46 | .52 | .36 | .31 | .39 | .22 | .32 | .23 | .21 | .16 |
| EU Leaders | .61 | .48 | .60 | .54 | .42 | .23 | .33 | .35 | .08 | .19 | .17 | .11 |
| Generic Users | .57 | .47 | .25 | .24 | .22 | .16 | .30 | .20 | .10 | .09 | .08 | .06 |

*Note.* (1) The signs for all the regression coefficients were positive. (2) Links with an "activating" hashtag have a hashtag in the first social tweet exchanged between the involved users.

**Figures**

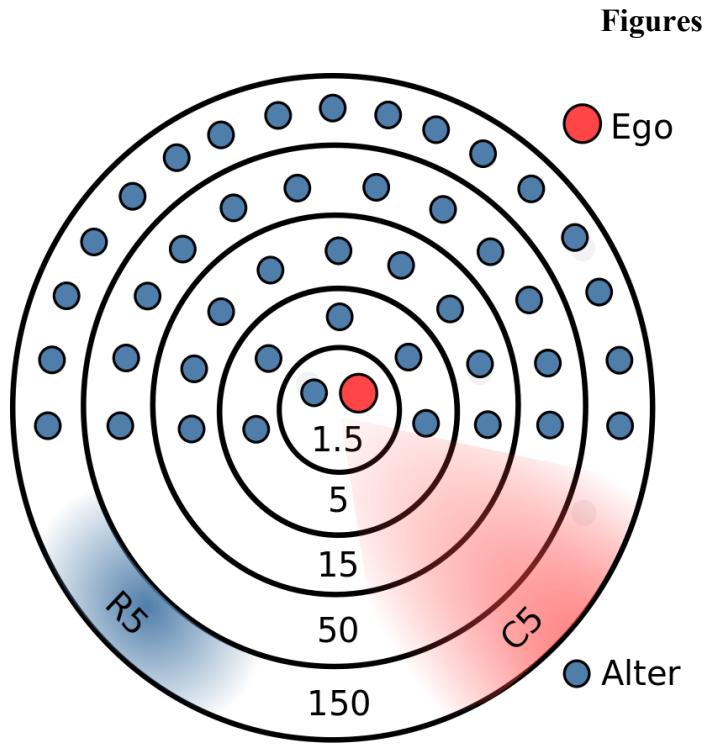

*Figure 1*. Ego network model. The *ego* is the focal individual of the model. *Alters* are social contacts of the ego. Rings (R1-R5, from the innermost one to the outermost one) are non-inclusive groups of alters with similar contact frequency. Circles (C1-C5) are inclusive groups. The typical hierarchical structure of human ego networks contains five circles with sizes averaging 1.5, 5, 15, 50, and 150.

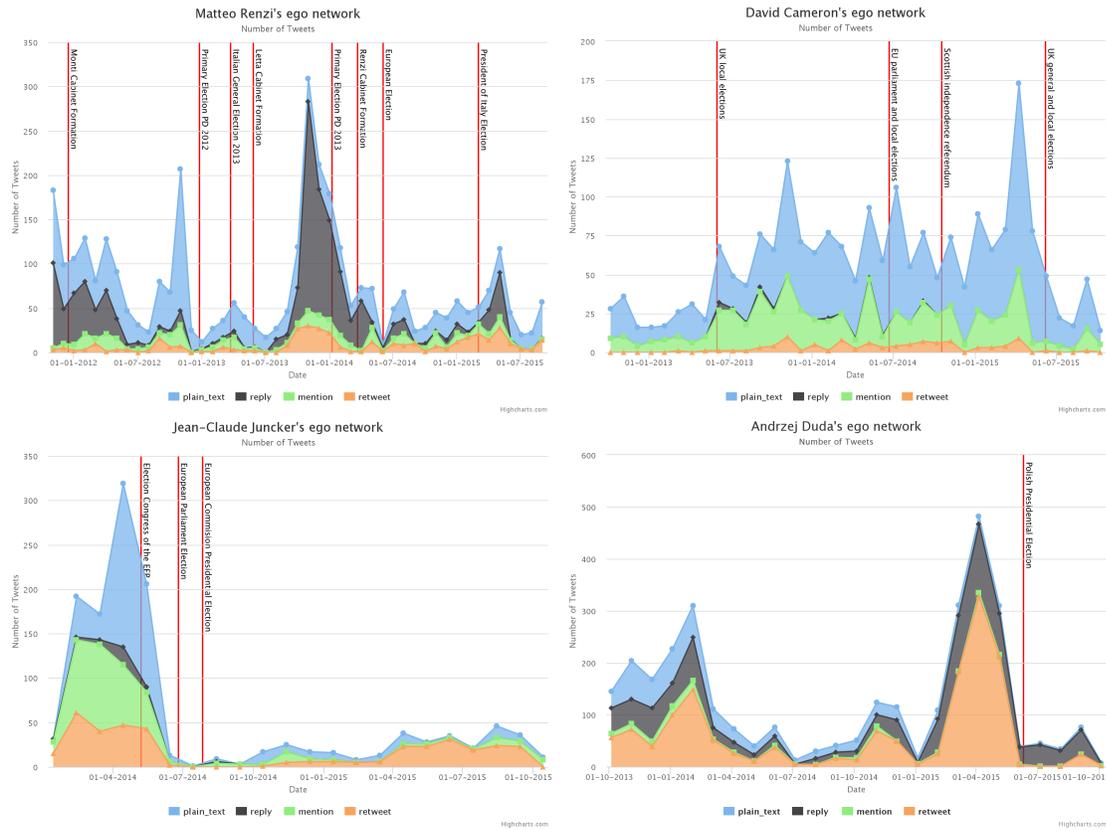

*Figure 2.* Number of plain and social tweets created per month by Matteo Renzi, David Cameron, Jean-Claude Juncker, and Andrzej Duda. Vertical lines in red represent the most important political events for these users between 2012 and 2015.

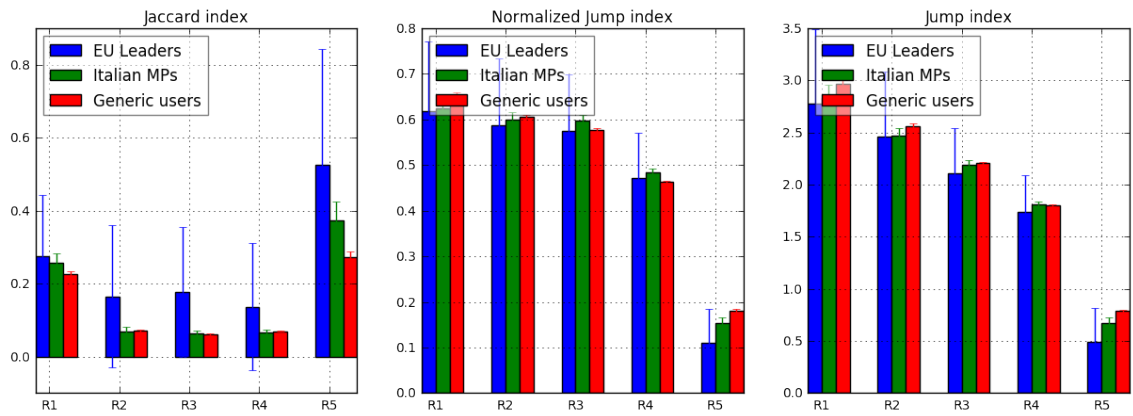

*Figure 3.* Stability (and instability) indices of ego network rings for EU Leaders, Italian MPs, and generic Twitter users. Jaccard index indicates the stability over time of each ring, whereas the Jump index indicates the variability of the rings.

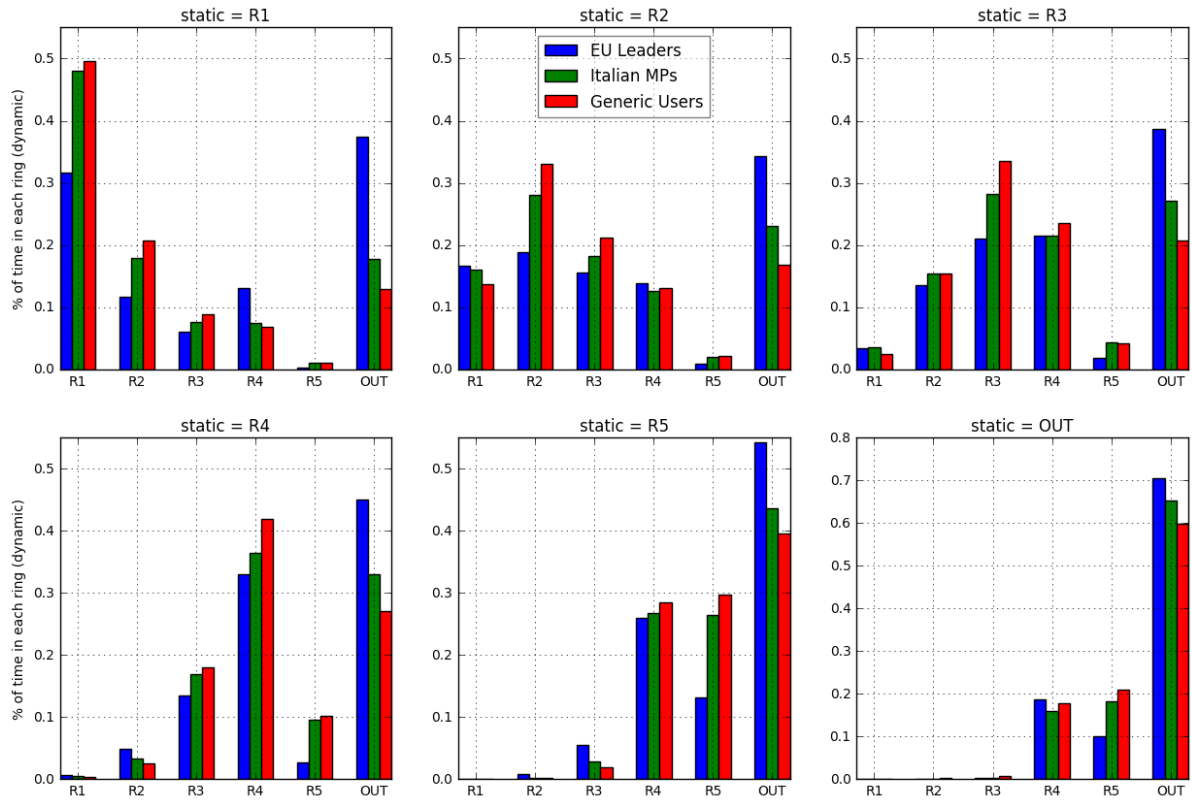

*Figure 4.* Relationship between the position within the ego network rings for relationships considering static and dynamic view of the ego networks.

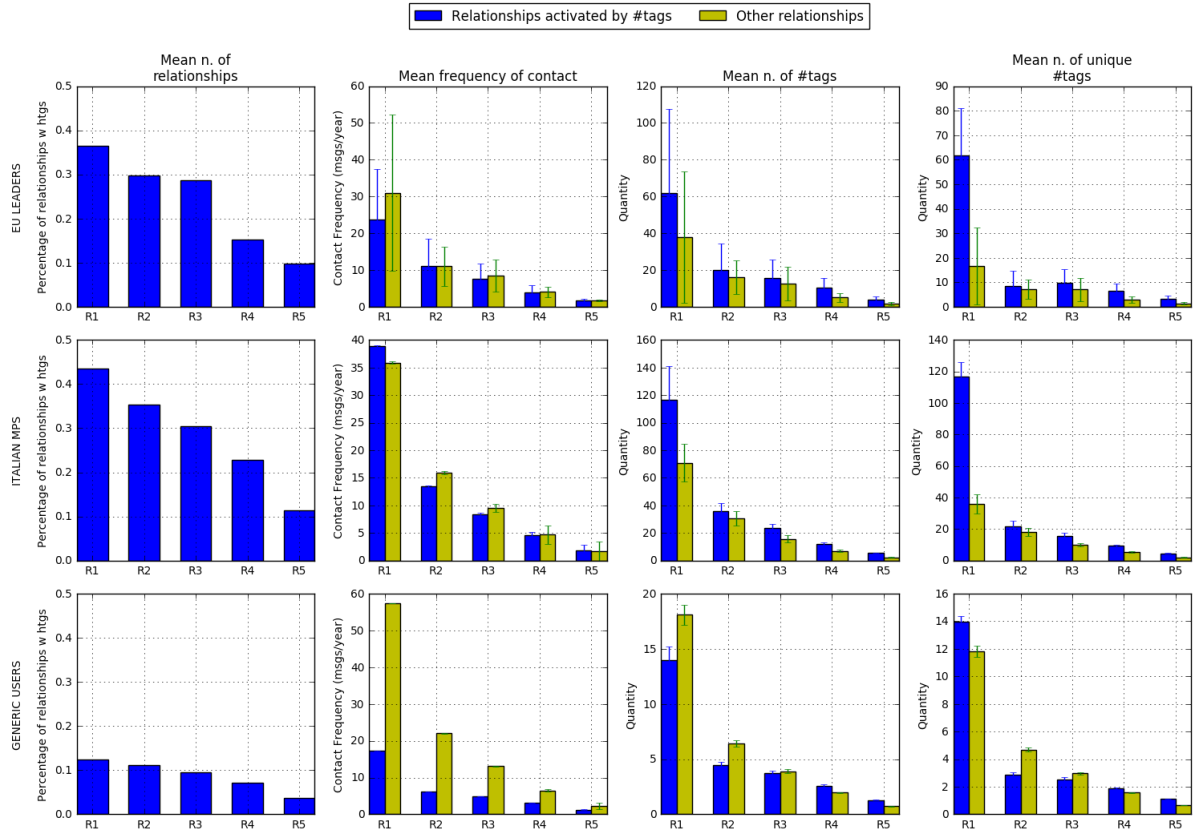

*Figure 5.* Statistics for social relationships activated by hashtags (#tags) and other relationships for EU Leaders, Italian MPs, and generic Twitter users, divided into the different ego network rings.